# Frequency stabilization of a quantum cascade laser by weak resonant feedback from a Fabry-Pérot cavity

GANG ZHAO,[1,2] JIANFEI TIAN,[3] JOSEPH T. HODGES,[1] AND ADAM J. FLEISHER[1,*]

[1]*Material Measurement Laboratory, National Institute of Standards and Technology, 100 Bureau Drive, Gaithersburg, MD 20899, USA*
[2]*Currently with the Institute of Laser Spectroscopy, State Key Laboratory of Quantum Optics and Quantum Optics Devices, Shanxi University, Taiyuan City 030006, Shanxi Province, China*
[3]*Institute of Laser Spectroscopy, State Key Laboratory of Quantum Optics and Quantum Optics Devices, Shanxi University, Taiyuan City 030006, Shanxi Province, China*
*\*Corresponding author: adam.fleisher@nist.gov*

**Frequency-stabilized mid-infrared lasers are valuable tools for precision molecular spectroscopy. However, their implementation remains limited by complicated stabilization schemes. Here we achieve optical self-locking of a quantum cascade laser to the resonant leak-out field of a highly mode-matched two-mirror cavity. The result is a simple approach to achieving ultra-pure frequencies from high-powered mid-infrared lasers. For short time scales (<0.1 ms), we report a linewidth reduction factor of $3 \times 10^{-6}$ to a linewidth of 12 Hz. Furthermore, we demonstrate two-photon cavity-enhanced absorption spectroscopy of an $N_2O$ overtone transition near a wavelength of 4.53 μm.**

The generation, measurement and dissemination of ultra-pure laser frequencies is at the forefront of emerging technologies in atomic, molecular and optical sciences. While many experiments operate at visible or near-infrared frequencies, emerging research in areas like cold chemistry [1] and molecular tests of the Standard Model [2] have pushed precision spectroscopy and optical coatings [3] into new frequency regimes. To probe molecular fingerprints using similarly ultra-pure frequencies, advances in the frequency stabilization of mid-infrared diode lasers like the quantum cascade laser (QCL) are required. For a recent review, see Consolino et al. [4].

Here we demonstrate a simple QCL frequency stabilization scheme which utilizes weak resonant optical feedback from a highly mode-matched Fabry-Pérot cavity with spherical mirrors. Contrary to all-electronic laser-stabilization schemes like the Pound-Drever-Hall method [5], optical self-locking to a reference cavity can readily suppress the high-frequency phase noise common to QCLs. Self-locking by optical feedback also yields a stronger linewidth reduction factor and enables cavity-enhanced sensing with high signal-to-noise ratios by easily locking to successive modes via laser frequency scanning. For textbook-level details, see Morville et al. [6].

Early demonstrations of diode laser frequency stabilization by resonant optical feedback used V-shaped or folded resonators to geometrically eliminate unwanted feedback from the direct cavity reflection [7-9]. The same general approach was later applied to the optical self-locking of a distributed feedback QCL [10]. Many publications reporting QCL locking to V-shaped cavities have followed, including a report of a QCL operating at a wavelength of 8.6 μm with a narrow 1-ms linewidth of 4 kHz [11].

Alternatively, two prior works have demonstrated diode laser self-locking to resonant optical feedback from a birefringent Fabry-Pérot cavity excited on-axis [12,13]. There, a quarter-wave plate and polarizing beam splitter were used to reject the direct cavity reflection, allowing only weak feedback from the birefringent cavity leak-out field to return to the laser. Salter et al. used transmission from a Fabry-Pérot cavity excited on-axis to seed diode laser self-locking via injection into the exit port of a Faraday isolator [14].

Neither birefringent cavities nor transmission feedback methods have been applied to QCLs. However, two papers have reported QCL self-locking to the leak-out field of a Fabry-Pérot cavity by mode-mismatching and then spatially filtering the direct reflection [15-17]. But intentional mode-mismatching is wasteful and greatly reduces the achievable intracavity power available for applications like optical trapping and nonlinear spectroscopy. Furthermore, when modeling competition between the spatially overlapped direct reflection and cavity leak-out fields, the relative phase of the two fields must be carefully considered. For these reasons, we propose the optical self-locking of a highly mode-matched QCL to a high-finesse Fabry-Pérot resonator excited on-axis.

We begin with a model for laser frequency stabilization by weak optical feedback from a two-mirror Fabry-Pérot cavity that is excited on-axis by the laser as illustrated in Fig. 1(a). We follow the general procedure [6,18-20] of replacing the laser exit facet power reflection coefficient with an effective coefficient accounting for all optical feedback fields. Then, in the limit of weak feedback, we calculate the coupled-laser frequency and gain at steady-state lasing conditions and compare to those of the free-running laser.

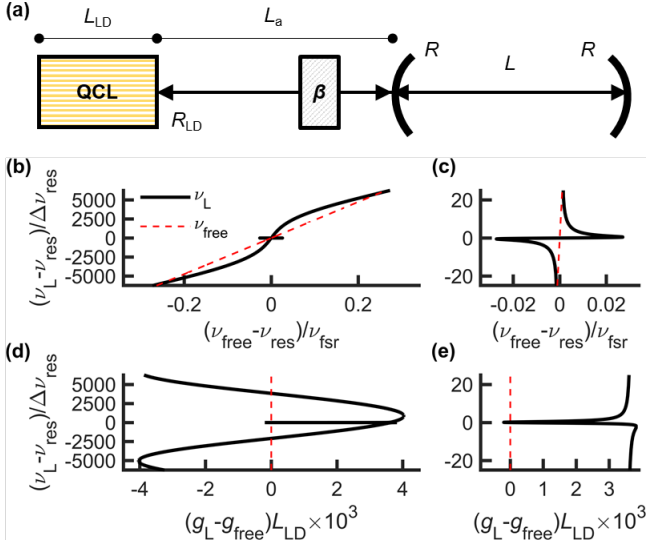

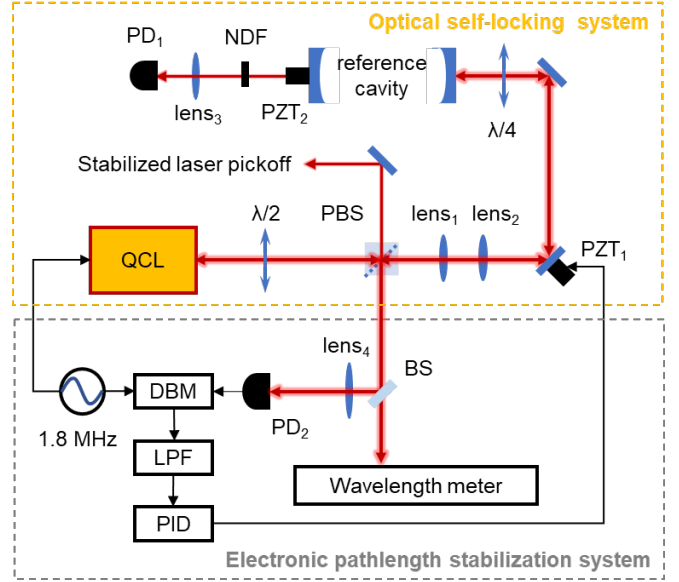

**Fig. 1.** Conceptualization (a) and model (b)-(e) for a quantum cascade laser (QCL) coupled to a Fabry-Pérot cavity by weak optical feedback. Coupled laser frequency detuning ($\nu_L - \nu_{res}$) relative to the reference cavity linewidth ($\Delta\nu_{res}$) is plotted versus (b) the free-running laser frequency detuning ($\nu_{free} - \nu_{res}$) relative to the reference cavity free spectral range ($\nu_{fsr} = \tau_{rt}^{-1}$) and (d) the normalized change in gain at threshold [($g_L - g_{free})L_{LD}$]. Panels (c) and (e) zoom in near resonance.

The reference cavity reflection function ($\tilde{h}_{OF}$) [6,19] can be written as the sum of a complex-valued leak-out term and a real-valued direct reflection component [21]:

$$\tilde{h}_{OF}(\omega) = \frac{t^2}{r}\frac{r^2\exp\{-i\omega\tau_{rt}\}}{1-r^2\exp\{-i\omega\tau_{rt}\}} - r \quad (1)$$

where $\omega = 2\pi\nu$ is the laser angular frequency, $t$ and $r$ are the reference cavity mirror electric-field transmission and reflection coefficients, respectively, $\tau_{rt} = 2L/c$ is the cavity round-trip time, $L$ is the single-pass cavity length, and $c$ is the speed of light.

Figure 1(b)-(e) illustrate QCL behavior near resonance as dictated by the reflection function in Eq. (1). In the weak feedback limit, the coupled laser angular frequency ($\omega_L$) and gain at threshold ($g_L$) as derived by Morville et al. [6,19] are:

$$\omega_{free} - \omega_L = \frac{c\sqrt{\beta(1+\alpha_H^2)}}{2n_{LD}L_{LD}}\frac{1-r_{LD}^2}{r_{LD}}\left[\text{Re}\{\tilde{h}_{OF}(\omega_L)\}\sin(\omega_L\tau_a + \theta) - \text{Im}\{\tilde{h}_{OF}(\omega_L)\}\cos(\omega_L\tau_a + \theta)\right] \quad (3)$$

$$g_L - g_{free} = \frac{\sqrt{\beta}}{L_{LD}}\frac{1-r_{LD}^2}{r_{LD}}\left[\text{Re}\{\tilde{h}_{OF}(\omega_L)\}\cos(\omega_L\tau_a) + \text{Im}\{\tilde{h}_{OF}(\omega_L)\}\sin(\omega_L\tau_a)\right] \quad (4)$$

where $\omega_{free}$ and $g_{free}$ are the free-running laser angular frequency and gain at threshold, respectively. In Fig. 1, we assume the special case where the power mode-matching factor is $\varepsilon = 1$, the reference cavity mirrors are lossless (i.e., $r^2 + t^2 = 1$) and the QCL-to-reference-cavity length ($L_a$) satisfies the general coupling condition $\omega_{res}\tau_a + \theta = 2\pi m$ where $m$ is an integer, $\tau_a = 2L_a/c$, $\theta = \tan^{-1}(\alpha_H)$, and $\alpha_H$ is the Henry factor [20]. The remaining model input parameters are: reference-cavity power reflection coefficient $R = r^2 = 0.99987$, cavity resonance wavelength $\lambda_{res} = 4.53$ μm, QCL exit facet power reflection coefficient $R_{LD} = r_{LD}^2 = 0.3$, laser gain-medium refractive index $n_{LD} = 3.3$, laser gain-medium length $L_{LD} = 3$ mm, $\alpha_H = 0.5$, and feedback power attenuation factor $\beta = 10^{-5}$.

**Fig. 2.** Experimental block diagram for QCL frequency stabilization. The orange dashed outline captures the optical (red arrows) self-locking system and the gray dashed outline captures the electronic (black arrows) pathlength stabilization system. Abbreviations: λ/2, half-wave plate; PBS, polarizing beamsplitter; PZT, piezo-electric transducer; λ/4, quarter-wave plate; NDF, neutral density filter; PD, photodetector; BS, beamsplitter; DBM, double-balanced mixer; LPF, loop filter; PID, proportional-integral-derivative servo.

The steady-state model suggests that frequency stabilization by resonant optical feedback from a Fabry-Pérot cavity is possible for a highly mode matched QCL. Figure 1 shows that the coupled-laser frequency is stabilized and the gain at threshold is reduced when the QCL is seeded by the leak-out field, even in the presence of direct reflection. The proper out-of-phase treatment of the feedback fields in Eq. (1) distinguishes the present model from that of a prior work in which the two fields were treated as in-phase [16]. To control the process from that point of view, Manfred et al. [16] relied upon coefficients to scale the relative amplitudes of the competing fields which were experimentally adjusted by mode mismatching.

A dynamic treatment of the QCL self-locking process using laser rate equations is beyond the scope of this work. However, we note that the reference cavity leak-out field will comprise both pumping and resonant frequencies during transient build-up [22], and therefore the reference cavity dynamics should also be considered.

We used the experimental apparatus illustrated in Fig. 2 to test our arguments. The reference cavity ($L = 0.75$ m) was formed by two high-reflectivity spherical mirrors with radius of curvature of 1 m inside of a vacuum enclosure comprising flexure mounts, vacuum viewports, and a central stainless-steel tube. We measured a cavity decay time constant of $\tau = 18.9$ μs, and therefore report the finesse, free spectral range, and cavity linewidth to be: $\mathcal{F} = 23\,700$, $\nu_{FSR} = 200$ MHz, and $\Delta\nu_{cav} = 8.44$ kHz (full-width at half maximum).

The distance between the QCL exit facet and the reference cavity was either $L_a \approx 2L$ or $L_a \approx L$. Active electronic stabilization of $L_a$ was achieved using a piezo-mounted mirror (PZT$_1$). An error signal was generated by modulating the QCL current driver at a frequency of 1.8 MHz and then demodulating the reference cavity reflection signal at photodetector PD$_2$. The resulting error signal was filtered, amplified, and sent to a servo controller with a 1 kHz bandwidth.

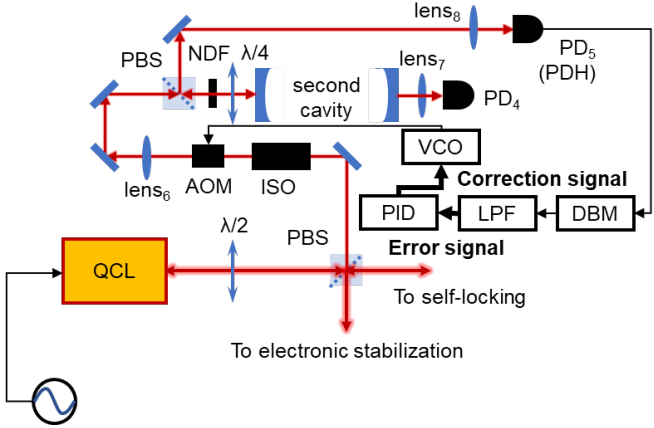

**Fig. 3.** Scheme for estimating the frequency-stabilized QCL linewidth. The error and correction signals used to Pound-Drever-Hall (PDH) lock the frequency stabilized QCL to a second high-finesse cavity of length $L_2$ = 1.5 m are identified by bold text and thick black arrows. Their sum provides a measurement of the total noise in the frequency stabilized QCL and second cavity system. New abbreviations: ISO, optical isolator; AOM, acousto-optic modulator; VCO, voltage-controlled oscillator.

Our set-up, without any tight aperture, allows for continuous tuning of the external optical feedback power attenuation factor $\beta_{\text{ext}}$ without reducing the optical power at the reference cavity. By adjusting a quarter-wave plate ($\lambda/4$) we controlled the power returning to the QCL exit facet after passing through the polarizing beam splitter (PBS) as depicted in Fig. 2. At its minimum value, from a combination of the PBS extinction ratio ($\beta_{\text{PBS}} < 2 \times 10^{-4}$) and the power mode-matching factor $\varepsilon \approx 0.5$, we estimate $\beta_{\text{ext}} < 1 \times 10^{-4}$.

To estimate the free-running QCL linewidth, we used a single-pass $N_2O$ sample cell and a Doppler-broadened molecular absorption feature with full-width at half maximum of 123 MHz as a frequency discriminator. To evaluate the frequency stabilized QCL, we used the measurement scheme shown in Fig 3. The locked-laser linewidth was estimated from the sum of the electronic error signal and correction signal sent to an acoustic-optic modulator (AOM) used to close a PDH locking loop, thus locking the frequency-stabilized QCL to a second high-finesse cavity of length $L_2$ = 1.5 m.

Frequency noise analysis is plotted in Fig. 4. For the free-running QCL (black trace), we estimate from the $\beta$-separation line [23] a linewidth of 4 MHz. For the stabilized QCL (red trace), the relatively flat portion of the power spectral density [24] of the PDH locking correction signal yields a fast 0.1-ms linewidth estimate of $\pi S_f(\phi)$ = 12 Hz. At Fourier frequencies less than 10 kHz, the locked laser linewidth is briefly subjected to $1/f^2$ noise and then presumably dominated by acoustic noises and vibrations of both external cavities as evidenced by the random walk behavior ($1/f^4$) [25]. For comparison, the in-loop noise from the optical feedback pathlength error signal (gray trace) yields a very low and flat power spectral density $\approx 10^{-4}$ Hz$^2$/Hz.

Theoretically, the coupled-laser linewidth is equal to the free-running linewidth scaled by the square of the reduced slope factor $p = d\omega_L/d\omega_{\text{free}}$, i.e., $\Delta\nu_{\text{locked}} = p^2 \Delta\nu_{\text{free}}$ [6,8,20]. We evaluated $p^2$ using Eqs. (1)-(3) and a refined value for the optical feedback power attenuation factor including the optical power overlap with the QCL facet, $\eta_{\text{LD}}$. Here, $\beta = \eta_{\text{LD}} \beta_{\text{ext}}$, where prior ray-tracing models suggest that $\eta_{\text{LD}} \ll 1$ for QCLs [16,26]. We also introduce mirror power losses, $\mathcal{L} = \ell^2$, which naturally adjusts the relative amplitudes of the two feedback fields in Eq (1): $r^2 + t^2 + \ell^2 = 1$. At $\eta_{\text{LD}}$ = 0.03 ($\beta \approx 3 \times 10^{-6}$) and $t^2 = \ell^2$, the reduced slope factor is $p^2 = 2.1 \times 10^{-6}$. This estimate is in good agreement with our experimental result of $p_{\text{exp}}^2 = 3 \times 10^{-6}$ at short timescales and predicts an optical self-locking range (e.g., Fig. 1(b)) of $(7.5 \times 10^{-3}) \nu_{\text{fsr}}$ = 1.5 MHz which is less than our QCL current modulation frequency of 1.8 MHz used to electronically stabilize $L_a$.

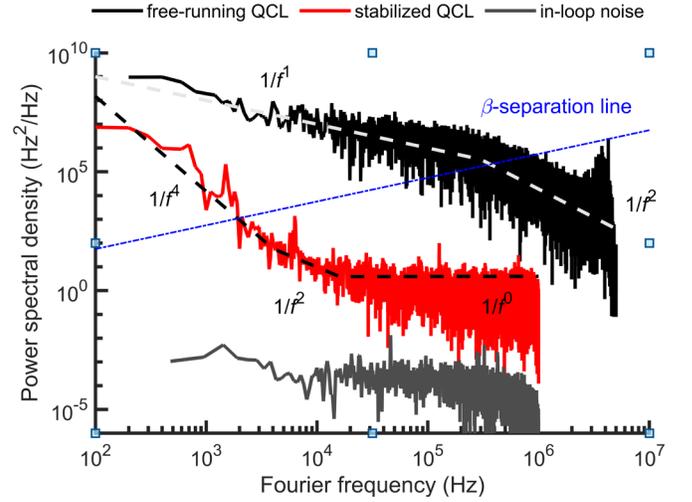

**Fig. 4.** Frequency noise analyses of the free-running and frequency-stabilized QCL. Plotted is the power noise spectral density for the free-running QCL (black), frequency-stabilized QCL (red) and in-loop signal relative to the reference cavity (dark gray). Also shown are the $\beta$-separation line to estimate the free-running QCL linewidth (blue dash-dotted line) and frequency noise models for the free-running QCL (light gray dashed line) the frequency-stabilized QCL (black dashed line).

Accessing the purity of the polarization state of the QCL and leak-out fields with high precision is a non-trivial task in the mid-infrared — and we acknowledge that we have not characterized the birefringence properties of our reference cavity (although it is likely similar to the birefringence previously measured by Fleisher et al. [27]). Regardless, we believe that the basic model (and experimental results) introduced here suggest that optical self-locking to the leak-out field of a linear two-mirror cavity is possible without needing to invoke mode-mismatching or birefringence. Any cavity birefringence would only act to increase the amplitude of the leak-out field relative to the direction reflection, and therefore improve the odds of optical self-locking relative to our baseline model. Very recently, optical self-locking of a multimode Fabry-Pérot QCL to a potentially birefringent Fabry-Pérot reference cavity excited on-axis was reported [3].

Finally, we performed nonlinear molecular spectroscopy with the frequency stabilized QCL. As noted earlier, our scheme enables high intracavity powers (≥10 W) while maintaining a low feedback power attenuation ratio using the quarter-wave plate and PBS cube. We introduced a gas sample of 20.4 μmol/mol $N_2O$ in air [28] into our reference cavity and tuned the locked laser frequency using PZT$_2$ across the $Q(18) \nu_3$-overtone two-photon absorption feature. In contrast to the two-photon cavity ring-down spectroscopy (TP-CRDS) of this transition that was previously reported [29], here we used the coupled-cavity laser transmission to perform two-photon cavity-enhanced absorption spectroscopy (TP-CEAS) [30].

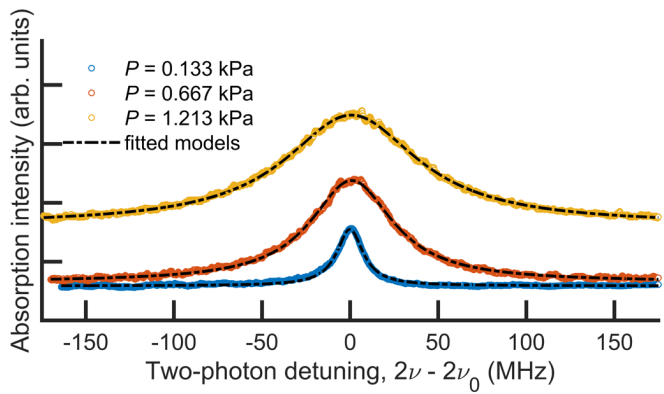

**Fig. 5.** Two-photon cavity-enhanced absorption spectroscopy of 20.4 µmol/mol N$_2$O in air [28] recorded at pressures ($P$): 0.133 kPa (blue), 0.667 kPa (red), and 1.213 kPa (orange). The resonant frequency of the N$_2$O $Q(18)$ $v_3$ overtone transition is $v_0$ = 66 179 400.8 MHz [30].

The TP-CEAS of N$_2$O recorded at three pressures ($P$) are plotted in Fig. 5. The observed Lorentzian two-photon half-widths at half maximum are: 9.0 MHz at $P$ = 0.133 kPa, 29 MHz at $P$ = 0.667 Pa, and 46 MHz at $P$ = 1.213 kPa. These values suggest that, in addition to pressure broadening [29], power broadening is also present. The relative frequency axis was approximated by prior calibration of the PZT$_2$ tuning against a commercial mid-infrared frequency comb system. In summary, our method of QCL self-locking to a high-finesse Fabry-Pérot cavity using concepts of polarization isolation [12] and high-bandwidth modulation [9] enables high intracavity powers and continuous tunability for sub-Doppler molecular spectroscopy.

Our self-locking scheme would be improved by the inclusion of a low thermal expansion cavity spacer and isolation chamber to reduce vibrational/acoustic noises and enable long-term stability (≥1 s). Alternatively, our length-tunable reference cavity could be further stabilized to a sub-Doppler molecular transition using slow electronic feedback to the piezo-electric transducer (PZT$_2$ in Fig. 2). With long-term stability achieved, we envision the direct stabilization of mid-infrared combs to reference QCL packages, forming robust mid-infrared metrology systems. With high throughput from the locked QCL, ultra-sensitive cavity ring-down spectroscopy becomes intriguing. For example, calculating the standard error [31] of the time constant fitted to ring-down data from the second cavity ($\tau$ = 33 µs) assuming a technical-noise limit and a maximum acquisition rate of $1/(8\tau)$, we project a hypothetically low noise-equivalent absorption coefficient of NEA ≈ $1 \times 10^{-11}$ cm$^{-1}$ Hz$^{-1/2}$.

**Funding.** National Institute of Standards and Technology (NIST), National Science Foundation of China (61875107, 61905136).

**Acknowledgments.** A. D. Ludlow (NIST) for discussing laser linewidth measurement methods; Q. Liu (NIST) for assisting in cavity assembly; C. E. Cecelski and J. Carney (NIST) for loaning a certified N$_2$O-in-air sample; K. K. Lehmann (University of Virginia) for discussing optical cavities and two-photon spectroscopy; and Z. D. Reed, R. W. Fox and D. M. Bailey (NIST) for commenting on the manuscript. Portions of this work were presented at the Conference on Lasers and Electro-Optics (CLEO) 2020, paper SF3G.2.

**Disclosures.** The authors declare no conflicts of interest.

**Data availability.** Data underlying the results presented in this paper will be made available online through NIST.

## References

1. J. L. Bohn, A. M. Rey, and J. Ye, Science **357,** 1002 (2017).
2. W. Ubachs, J. C. J. Koelemeij, K. S. E. Eikema, and E. J. Salumbides, J. Mol. Spectrosc. **320,** 1 (2016).
3. G. Winkler, L. W. Perner, G.-W. Truong, G. Zhao, D. Bachmann, A. S. Mayer, J. Fellinger, D. Follman, P. Heu, C. Deutsch, D. M. Bailey, H. Peelaers, S. Puchegger, A. J. Fleisher, G. D. Cole, and O. H. Heckl, Optica doi:10.1364/OPTICA.405938. (For review only: arXiv:2009.04721v1 (2020).)
4. L. Consolino, F. Cappelli, M. S. de Cumis, and P. De Natale, Nanophotonics **8,** 181 (2019).
5. R. W. P. Drever, J. L. Hall, F. V. Kowalski, J. Hough, G. M. Ford, A. J. Munley, and H. Ward, Appl. Phys. B **31,** 97 (1983).
6. J. Morville, D. Romanini, and E. Kerstel, "Cavity enhanced absorption spectroscopy with optical feedback," in *Cavity-enhanced spectroscopy and sensing*, G. Gagliardi and H.-P. Loock, eds. (Springer, 2014), pp. 163-209.
7. B. Dahmani, L. Hollberg, and R. Drullinger, Opt. Lett. **12,** 876 (1987).
8. Ph. Laurent, A. Clairon, and Ch. Bréant, IEEE J. Quantum Electron. **25,** 1131 (1989).
9. A. Hemmerich, D. H. McIntyre, D. Schropp, Jr., D. Meschede, and T. W. Hänsch, Opt. Commun. **75,** 118 (1990).
10. G. Maisons, P. Gorrotxategi Carbajo, M. Carras, and D. Romanini, Opt. Lett. **35,** 3607 (2010).
11. E. Fasci, N. Coluccelli, M. Cassinerio, A. Gambetta, L. Hilico, L. Gianfrani, P. Laporta, A. Castrillo, and G. Galzerano, Opt. Lett. **39,** 4946 (2014).
12. C. E. Tanner, B. P. Masterson, and C. E. Wieman, Opt. Lett. **13,** 357 (1988).
13. J. Morville and D. Romanini, Appl. Phys. B **74,** 495 (2002).
14. R. Salter, J. Chu, and M. Hippler, Analyst **20,** 4669 (2012).
15. A. G. V. Bergin, G. Hancock, G. A. D. Ritchie, and D. Weidmann, Opt. Lett. **38,** 2475 (2013).
16. K. M. Manfred, L. Ciaffoni, and G. A. D. Ritchie, Appl. Phys. B **120,** 329 (2015).
17. D. A. King and R. J. Pittaro, Opt. Lett. **23,** 774 (1998).
18. R. Lang and K. Kobayashi, IEEE J. Quantum Electron. **16,** 347-355 (1980).
19. J. Morville, S. Kassi, M. Chenevier, and D. Romanini, Appl. Phys. B **80,** 1027 (2005).
20. J. Ohtsubo, *Semiconductor Lasers: stability, instability and chaos*, 3rd ed. (Springer, 2013).
21. A. E. Siegman, *Lasers* (University Science Books, 1986).
22. A. Cygan, A. J. Fleisher, R. Ciuryło, K. A. Gillis, J. T. Hodges, and D. Lisak, Comms. Phys. **4,** 14 (2014).
23. G. Di Domenico, S. Schilt, and P. Thomann, Appl. Opt. **49,** 4801 (2010).
24. L. Hollberg, V. L. Velichansky, C. S. Weimer, and R. W. Fox, "High-accuracy spectroscopy with semiconductor lasers: application to laser-frequency stabilization," in *Frequency control of semiconductor lasers*, M. Ohtsu, ed. (Wiley, 1996), pp. 73-93.
25. D. A. Howe, D. W. Allan, and J. A. Barnes, "Properties of signal sources and measurement methods," in *Characterization of clocks and oscillators: Technical Note 1337*, D. B. Sullivan, D. W. Allan, D. A. Howe, and F. L. Walls, eds (National Institute of Standards and Technology, 1990), pp. TN14-TN60.
26. K. M. Manfred, "Mid-infrared laser spectroscopy for trace gas detection," Ph.D. dissertation (University of Oxford, 2015), pp. 165-166.
27. A. J. Fleisher, D. A. Long, Q. Liu, and J. T. Hodges, Phys. Rev. A **93,** 013833 (2016).
28. M. E. Kelley, G. C. Rhoderick, and F. R. Guenther, Anal. Chem. **86,** 4544 (2014).
29. G. Zhao, D. M. Bailey, A. J. Fleisher, J. T. Hodges, and K. K. Lehmann, Phys. Rev. A **101,** 062509 (2020).
30. K. K. Lehmann, J. Opt. Soc. Am. B **37,** 3055 (2020).
31. H. Huang and K. K. Lehmann, J. Phys. Chem. A **117,** 13399 (2013).